\documentclass[aps,pra,twocolumn,amsfonts,amsmath,amssymb,superscriptaddress]{revtex4-1}%
\pdfoutput=1
\usepackage{amsfonts}
\usepackage{color}
\usepackage{amsmath}
\usepackage{amssymb}
\usepackage{graphicx}%
\begin{document}
\title{Classical simulation of boson sampling with sparse output}

\date{\today}
\author{
Wojciech Roga 
}
\author{
Masahiro Takeoka
}
\address{
National Institute of Information and Communications Technology, 
Koganei, Tokyo 184-8795, Japan\\ E-mail: wojciech.roga@nict.go.jp
}

\begin{abstract}
Boson sampling can simulate physical problems for which classical simulations are inefficient. However, not all problems simulated by boson sampling are classically intractable. We show explicit methods of classically simulating boson sampling when its outcome is known to be highly sparse. In the methods, we first determine a few marginal distributions and then recover the joint distribution from them. Although the latter could be of high complexity in general, we show that it can be classically calculable when the high sparsity assumption holds. Various extensions are discussed including a version involving quantum annealing.
\end{abstract}

\maketitle
{\it Background and motivation.--} Simulating complicated quantum systems on classical or quantum simulators is an interesting problem with the industrial impact. It is simply cheaper to test many, e.g., molecular configurations in the simulators than synthetizing the molecules and testing their crucial properties experimentally. However, some problems are believed to be of complexity for which classical computers are inefficient. For instance, Huh et al. \cite{Huh,Huh2} showed that the statistics of Franck-Condon (FC) factors \cite{Franck,Condon} for vibronic transitions in large molecules \cite{Doktorov,Jankowiak} is equivalent to the statistics of samples in a version of boson sampling  \cite{Aronson,Spring,Tillman,Shchesnovich,Tichy,Motes,Gard} - the Gaussian boson sampling \cite{Lund,Hamilton,Quesada}. Although, it is widely accepted that boson sampling from interferometers described by the average-case Haar-random unitary transformations or Gaussian-random matrices is classically computationally inefficient \cite{Aronson}, it is not clear if particular problems of quantum chemistry belong to this class, as the related matrices are not typically Haar- or Gaussian-random \cite{Huh,Cao18}. In particular, if these matrices were of low rank or consisted of non-negative numbers there exist efficient algorithms for approximating their permanents \cite{Barvinok,Jerrum}. This reduces complexity of scattershot boson sampling. 

In this paper we discuss the case when we {\it a priori} know that the statistics of outputs from boson sampling is sparse. This knowledge can be based on experience with similar problems, symmetries or other physical properties. We analyze examples of relevant physical systems with approximately sparse vibronic-spectra that can be simulated on a Gaussian boson sampling-type simulator in \cite{Jacob}. There, sparsity of a spectrum is expected based on shapes of spectra of typical molecules and the intuition that the most significant transitions in 0K are likely these with only a few phonons involved in just a few modes. In the present paper, we introduce methods of at least approximate classical computation of the sparse statistics. We use the fact that for boson sampling marginal distributions for small number of modes can be efficiently calculated. Then we apply appropriately modified compressive sensing methods to efficiently recover the joint statistics. Up to our knownledge this is the first result dealing with classical simulability of boson sampling with sparse output. Moreover, we develop original methods to improve the efficiency of the classical algorithms for compressive sensing reconstruction, namely the so-called polynomial time matching pursuit (PTMP) that can be extended to other methods, for instance, gradient pursuit \cite{Blumensath08}. Finally, our novel approach has an impact on interdisciplinary studies on molecular vibronic spectroscopy  \cite{Jacob}, compressive sensing, and quantum computing with hybrid devices.  

Let us summarize our arguments. For scattershot boson sampling computability of marginal distributions for small number of modes is implied by the result of Gurvits cited together with the proof in the Aaronson and Arkhipov paper \cite{Aronson} as follows: 

{\bf Theorem (Gurvits’s k-Photon Marginal Algorithm)}{\it There exists a deterministic classical algorithm that, given a unitary matrix $U\in \mathbb{C}_{M\times M}$, indices $i_1, . . . , i_k \in[M]$, and occupation numbers $j_1, . . . , j_k\in\{0, . . . , N\}$, computes the joint probability
$$Pr_{S=(s_1,...,s_M)\sim D_U}[s_{i_1} = j_1 \wedge ... \wedge s_{i_k} = j_k]$$
in $N^{O(k)}$ time. }\\

\noindent Here, $S=(s_1,...,s_M)\sim D_U$ means that the occupation numbers $S$ are sampled according to the probability distribution over possible outputs of $U$. If $k$ is small, as we assume in this paper, calculating marginal distributions is efficient. The counterpart of this theorem for Gaussian boson sampling is discussed in the conclusions. In our approach we use the compressive sensing methods \cite{Donoho,Candes1,Baraniuk,Candes2,Foucard,Draganic,Candes3} to recover the joint sparse distribution from marginal ones. We show how to do that efficiently. Our arguments are inspired by works from the field of quantum compressive sensing \cite{Cramer,Gross,Liu,Flammia,Shabani,Shabani2} or similar ones that consider recovering full information about states of high dimensional systems from states of low-dimensional subsystems. In particular, Cramer et al. \cite{Cramer} proposed an algorithm to reconstruct a low rank quantum state from its reduced density matrices known from quantum tomography of subsystems. They used a modified singular value thresholding algorithm (MSVT). The essential part of it is finding the ground states of matrices that can be thought as Hamiltonians. This operation itself is inefficient, but if the Hamiltonian is constructed from local nearest-neighbor interactions the best so-called matrix product state approximations of the ground states can be found efficiently. Then MSVT converges to the right solution in time polynomial in the size of the system. Additionally, the application of the matrix product states representation allows for significant reduction of the required memory.  The procedure we provide is similar to this idea. We show that the complexity bottleneck of a compressive sensing reconstruction of a probability vector from marginal distributions is in localizing the largest element from a long list (counterpart of finding the ground state). This procedure typically would require the number of steps and bits of memory which is $O(d)$, where $d$ is the length of the list. However, we show that in our case the list may be written as the local nearest-neighbor interaction Hamiltionian of a classical spin chain. Localization of the maximum energy configuration is efficient in this problem both in terms of the number of operations and memory \cite{Ising,Schuch10}.

Our work is related to a general problem whether sparsity of the output of a quantum computer guarantees classical simulability. In \cite{Schwarz,Pashayam} there was shown that the existence of an efficient classical randomized algorithm for entries of all marginal distributions and the sparsity assumption indeed guarantee this for the circuit-based computing device. However, this result is not automatically applicable to boson sampling. Moreover, in our approach we relax one of the assumptions considering only computability of some marginal distributions for limited number of modes which extends the class of known classically tractable problems consistent with the boson sampling architecture.

\begin{figure}[h]
\center
\includegraphics[width=0.45\textwidth]{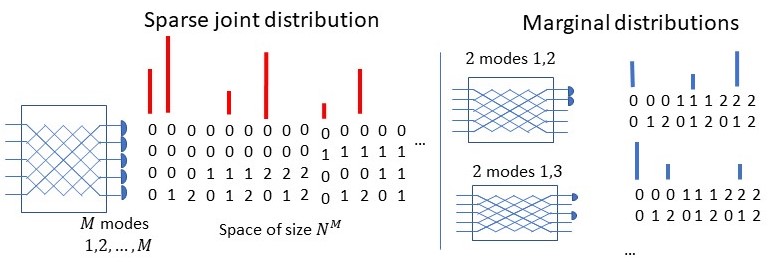}
\caption{Red or blue bars - the statistics of simultaneous readouts of different numbers of photons at the output of an interferometer. Left hand side: The distributions measured directly by $M$ photon-number-resolving detectors. All combinations of possible numbers of photons in $M$ modes form an $N^M$ dimensional vector. Right hand side: Marginal distributions of the occupation numbers in two chosen modes.}
\label{diag1}
\end{figure}

{\it Marginal distributions.---} Let us start considering the following scenario. For some unitary transformation $U$ which reflects features of a physical system we want to simulate the transition probabilities $x=|\langle \Psi|U|n_1,n_2,...,n_M\rangle|^2$ from a given $M$-mode initial state $|\Psi\rangle$ to all occupation numbers states  $|n_1,n_2,...,n_M\rangle$ (the left hand side of fig. \ref{diag1}) that may be recorded by simultaneous readouts of $M$ phtoton resolving detectors. We assume that there can be $0,1,...,N-1$ possible photons in each output mode. So, the searched vector $x$ is of length $N^M$. Here, we allow for the total photon number not being preserved. In this way we can consider the situation with losses or Gaussian boson sampling using the same formalism.
We claim that if $x$ is sparse, i.e., it contains at most $\mathfrak s$ non-zero entries, where $\mathfrak s\ll N^M$, it is possible to efficiently find $x$ calculating the statistics of marginal distributions from smaller number of detectors measuring only chosen modes and ignoring the rest (the right hand part of fig. \ref{diag1}). 

To analyze the problem in detail let us introduce the following notation. Vector $x$ can be decomposed in the basis of measured sets of occupation numbers as follows
\begin{equation}
x=\sum_{n_1,...,n_M}\alpha_{n_1,...,n_M}|n_1,...,n_M).
\end{equation}
Here, the numbers $\alpha_{n_1,...,n_N}$ are non-negative, sum up to one and only $s$ of them are non-zero. 
We will use the following convention
\begin{equation}
\begin{array}{c c c c c c}
|n_1,n_2,...)^T=&(\cdot\cdot\cdot 1 \cdot \cdot)&\otimes&(\cdot\cdot 1 \cdot\cdot\cdot)&\otimes&...\\
\ &\uparrow&\ &\uparrow& \ \\
\ &n_1+1&\ &n_2+1 &\ &...
\end{array}
\label{tensorrepres}
\end{equation}
where the lower line indicates the positions of 1 in each vector component of the tensor product and there is appropriate number of 0s in place of dots. In this explanation we will use a notation for two-detector simultaneous readouts, however 
the formalism can be extended to simultaneous readouts of a different number of detectors if necessary. The measurement of modes $i$ and $j$ leads to the marginal probability distribution 
\begin{equation}
y_{n_i, n_j}=A_{n_i,n_j}x
\label{ope}
\end{equation}
where $y_{n_i, n_j}$ is the sum of all entries $\alpha_{n_1,...,n_N}$ with fixed $n_i$ and $n_j$. We get this distribution observing frequencies of outcomes from given modes independently of what happens in the remaining modes. 
In the chosen convention the rows of the so-called {\it measurement matrix} $A$ are binary patterns as follows
\begin{equation}
A_{n_i,n_j}=\gamma_{n_i}\odot \gamma_{n_j},
\label{gammagamma}
\end{equation}
where $\odot$ means the entry-wise multiplication and  
\begin{equation}
\begin{array}{c c c c c c}
\gamma_{n_i}=&{\bf 1}^{\otimes i-1}&\otimes&(\cdot\cdot 1 \cdot\cdot \cdot)&\otimes&{\bf 1}^{\otimes M-i}\\
\ &\ &\ &\uparrow& \ & \\
\ &\ &\ &n_i+1 &\ &
\end{array}.
\label{gammam0}
\end{equation}
Here ${\bf 1}=(1,1,1,...)$ and $(\cdot\cdot 1\cdot\cdot \cdot)$ are $N$ dimensional vectors. The entry-wise multiplication preserves the tensor product structure and can be executed as the entrywise multiplication in each part of the tensor product.

Probabilities of $k$ different simultaneous readouts form a $k$-dim vector 
\begin{equation}
 y=Ax,
\end{equation}
where $A$ is a $k\times N^M$ binary matrix.  If $x$ is sparse in the basis incoherent with the rows of the measurement matrix it can be determined based on the number of measurements $k\ll N^M$ as the most sparse solution of $y=Ax'$. This can be seen as the constraint $l1$ norm minimization problem. It is solvable by many known algorithms used in compressive sensing \cite{Draganic}. In the next part we analyze the complexity of particular algorithms adapted to minimize the computational costs and memory requirements.

{\it Complexity analysis.--} Assume that we are focused on simultaneous readouts of different photon numbers in neighboring modes. The corresponding marginal distributions can be calculated in polynomial time in the number of modes. Indeed, Gurvits’s Algorithm from \cite{Aronson} calculates $k=N^2M$ of coincidences in all neighboring modes in $N^{O(2)}N^2M$ steps. The dimensionality of the sparse vector is $d=N^M$. We allow for sub-linear scaling of the number of non-zero entries $\mathfrak s$. Therefore, we keep $O(\mathfrak s)$ and $O(d)$ well separated. Moreover, we assume that problems of complexity $O(\mathfrak s)=O(k)$ are efficiently tractable. From the readouts for marginal distributions, we want to recover the most sparse joint distribution knowing the measurement matrix. So, our goal is to solve the underdetermined problem $y=Ax$, where $y$ is a $k$-dimensional measurement vector of marginal distributions, $x$ is a $d$-dimensional sparse vector which is searched.  In our case, rows of $A$ are well structured patterns. This implies that it is easy to multiply $A$ by any sparse vector. For instance, assume that we know that an entry $\kappa$ of a vector is non-zero. We decompose $\kappa$ as a N-inary number which gives us immediately its tensor product representation as in (\ref{tensorrepres}) consistent with the structure of $A$. For example, assuming that $N=2$, position $14$ is represented as $01101$ (corresponding to binary 13 as 0 occupies the first position) which corresponds to
\begin{equation}
(1 0)\otimes (0 1)\otimes (0 1)\otimes (1 0)\otimes (0 1).
\end{equation}  
This vector has just one non-zero element in the 14th entry. It is immediate to show what is its overlap with a particular row of $A$ which, for instance, can be of the form
\begin{equation}
(1 1)\otimes (0 1)\otimes (1 0)\otimes (1 1)\otimes (1 1).
\end{equation}  
We need to check only relevant modes.

To solve the underdetermined problem $y=Ax$ we discuss in detail two first-order greedy algorithms. We consider the matching pursuit \cite{mallat} method which is simple but in general less accurate, and the gradient pursuit which can be more accurate and faster but slightly more computationally demanding \cite{Blumensath08}. These two algorithms are enough to discuss the bottleneck for the memory and computational costs, and to show how to overcome these problems. In particular, our modification of the matching pursuit leads to a new algorithm which we call the polynomial time matching pursuit (PTMP).

The standard matching pursuit protocol finds the $s$-sparse solution. It is summarized as follows:
I. (Initialization) At step $0$: the residual $r^0=y$ and the approximate solution is $x^0=0$. 
II. (Support detection) In step $i$ recognize the column of $A$ denoted by $A^t$ which is the most similar to the current residue $r^{i-1}$ by solving $t={\rm argmax}_{t'}|(A^T r^{i-1})_{t'}|$. 
III. (Updating) Update the solution only in index $t$ i.e., $x^i_t=x^{i-1}_t+(A^T r^{i-1})_t$ and update the residue using $t$-th column of $A$ as follows $r^i=r^{i-1}-(A^T r^{i-1})_tA^t$.
IV. Continue iterating until a stopping criterion is matched.

Let us notice that the first and the third part of the algorithm can strongly benefit from the sparsity of vectors involved. Moreover, for any $t$, vector $A^t$ can be found operationally (multiplication of $A$ and a sparse vector). So, $A^t$ does not need to be stored beforehand and the memory and computational costs of these parts are $O(k)$ - size of $r_i$. The entire procedure can be iterated until a given sparsity $\mathfrak s$ of the solution is achieved. 

Finally, let us consider the operational costs of the support detection part. Without exploiting the structure of $A$ we require $O(kd)$ bits of memory to store the matrix and the same for the operational costs for multiplication  $A^T r_{i-1}$. Moreover, without smart tricks, typically, we would need at least $O(d)$ steps to find the maximum value from the list  $A^T r_{i-1}$. The same amount of memory is needed to store the list. However, considering specific features of the problem we can overcome the bottleneck. As for finding the index of the largest element of a list, we notice that it is equivalent to finding the leading eigenvector of the diagonal matrix (Hamiltonian) with the list on the diagonal. We know that for at least local Hamiltonians there are computationally efficient methods for finding the eigenvectors. 
Let us notice that in our problem $A^T r$ in the support detection part can be written exactly as a diagonal local Hamiltonian. 
To simplify the explanation let us consider first an example with the readouts from a single detector only. A row of measurement matrix $A$ corresponding to $n_m$ photons in mode $m$ is given as $\gamma_{n_m}$ in (\ref{gammam0}). Observing all $n_m$ from only mode $m$ we have
\begin{equation}
A_{[m]}^T r_{[m]}=\left({\bf 1}^{\otimes m-1}\otimes(r_{0_m},r_{1_m},r_{2_m},...)\otimes {\bf 1}^{\otimes M-m}\right)^T,
\end{equation}
where $A_{[m]}$ is a submatrix of $A$ corresponding to different photon numbers recorded in mode $m$. Here, $r_{[m]}$ is part of the residual vector that corresponds to rows of $A_{[m]}$. Measurements of other modes have also form of the local Hamiltonians if understood as diagonal matrices. So, the same holds for $A^T r$. For simultaneous readouts of $2$ neighboring modes we have 
\begin{eqnarray}
&&A_{[m,m+1]}^T r_{[m,m+1]}=\label{goodness}\\ 
&&\left({\bf 1}^{\otimes m-1}\otimes(r_{0_m,0_{m+1}}\!,r_{0_m,1_{m+1}}\!,r_{0_m,2_{m+1}},\!...)\otimes {\bf 1}^{\otimes M-m-1}\right)^T,\nonumber
\end{eqnarray}
which written in the diagonal form is a part of a typical nearest neighbor local Hamiltonian for a one dimensional spin chain. It is clear from this notation that the matrix vector multiplication $A^T r$ requires negligible operational costs and $O(k)$ bits of memory to store $d$ long vector in its compressed representation as a sum of local matrices.

Using these observations, the support finding from the matching pursuit algorithm can be determined much faster than in $O(d)$ time. If we consider just two neighboring modes simultaneous readouts our problem can be described in terms of the classical spin chain formalism. We can use an explicit strategy from, e.g., \cite{Schuch10} to find the optimal configuration of the classical spin chain which is equivalent to finding the position of the maximal value in our $A^Tr$ list. Indeed, $h_{m,m+1}(i_m,i_{m+1})$ from the algorithm \cite{Schuch10} is equivalent to $r_{(i-1)_m,(i-1)_{m+1}}$. This procedure reduces the computational costs of the support detection to $2MN^2$ (factor 2 is from the necessity to repeat the procedure for $-A^Tr$ as we are looking for the largest element in the absolute value). The matching pursuit with the modification increasing its efficiency is called the polynomial time matching pursuit (PTMP). We use this method to reconstruct vibronic spectra of some molecules from their marginal distributions in \cite{Jacob}.

PTMP although computationally simple is not the fastest one as the same support index may need to be used many times. Also it does not guarantee the accurate solution, as the solution is expressed based on relatively small number of columns of $A$ in which $y$ is decomposed. Some modifications of the method can lead faster to more accurate results. Among them there is an orthogonal matching pursuit algorithm (OMP) \cite{zhang} in which in each step the support is updated and kept in the memory. The solution is approximated by the vector of coefficients in the best 2-norm approximation of the measurement vector $y$ in terms of the selected columns of $A$. Another algorithm, gradient pursuit (GP)  \cite{Blumensath08} updates the solution by correcting it in the direction of the largest gradient of the 2-norm mentioned above by a given step. As discussed in \cite{Blumensath08} the only additional cost over what we have in the matching pursuit is the cost of calculating the step size. This can be however done efficiently in PTMP due to an easy way of finding the product of a sparse vector and matrix $A$. The convergence rates depend on contracting properties of $A$ and a chosen algorithm. 

{\it Accuracy of the reconstruction.--} For successful compressive sensing reconstruction the so-called coherence between the rows of a measurement matrix and the sparsity basis must by low. Roughly speaking it means that a particular measurement is sensitive to changes in a large part of the measured vector. For this reason random unstructured matrices are of particular interest as they are incoherent with almost any basis. However, random matrices are problematic for large scale tasks. They are expensive in terms of storage and operational costs as the inefficient matrix vector multiplication is usually needed \cite{Cevher}. Therefore, structured matrices which can be defined operationally and do not need to be stored are also desired. In our case, measurement matrix $A$ is structured and of low coherence with respect to the measured space basis. However, the usefulness of a specific matrix depends as well on the reconstruction algorithm. As we do not have theoretical predictions regarding convergence of considered protocols with the measurement matrix, we tested the performance of the GP algorithm with $A$ numerically. We have measured 1000 randomly chosen distributions with $\mathfrak s=4,5,6$ non-zero entries for the problem with $M=6$ output modes and $N=4$ different events measured in each mode. Matrix $A$ was associated with all 2-neighboring modes coincidence measurements. So, $A$ is a $80\times 4096$ matrix. We have tested how often $X=x_{test}^Tx_{solved}/|x_{test}|^2$  is larger than a given threshold, where $x_{test}$ and $x_{solved}$ are the randomly chosen measured signal used in the simulation and the reconstructed signal respectively. We iterated the algorithm not more than 50 times. For $\mathfrak s=4$ we observe that in about $80\%$ cases $X>0.9$ and in $74\%$ cases $X>0.99$. For $\mathfrak s=5$, we have $X>0.9$ and $X>0.99$  in $64\%$ and $56\%$ of situations respectively. Finally, for $\mathfrak s=6$ we observe $X>0.9$ and $X>0.99$ in $47\%$ and $37\%$ of cases respectively. As there are many possible more complicated and more accurate algorithms which can still share the feature of the reduced complexity we predict that these numbers can be improved. However, testing them further is out of the scope of this paper.  How PTMP works in practice is analyzed in a separate paper \cite{Jacob}. There,
 we observe that the distribution can be only approximately sparse fort the algorithm to reconstruct the most significant picks.

{\it Conclusions.--} In this paper we show that if we {\em a priori} know that boson sampling samples according to a sparse distribution and the sparsity is high enough we can calculate the first-order approximation of this distribution efficiently on a classical computer. The crucial steps are to calculate the marginal distributions using the Gurvits’s k-Photon Marginal Algorithm and then to use an algorithm for compressively sensed sparse signal recovery. Many of these algorithms quickly converge assuming that the  support localization, as discussed in this paper, is efficient. We show that due to the specific form of the marginal measurements the problem can be reduced to finding the optimal configuration in a 1 dimensional classical spin chain with local interactions. The latter is known to be efficient.

For Gaussian boson sampling the only difference is in computing the marginal distributions. In this case evolving an input states through the interferometer and finding partial states for given modes is easy \cite{Barlett}. The statistics of photons of these states requires computing loop Hafnians \cite{Hamilton, Quesada,Kruse,Quesada2} of appropriate matrices which for two mode Gaussian states are still classically tractable. In consequence, we can use the relation between Franck-Condon factors and the Gaussian boson sampling \cite{Huh} and apply the algorithm described in this paper to find Franck-Condon factors under the condition that their distribution is sparse. This new approach is tested in \cite{Jacob}.     

In this paper we have investigated the situation with only nearest neighbor modes measurements. This restricts the tolerable sparsity for the method. When the sparsity decreases ($\mathfrak s$ increases) a larger amount of data is needed. Implementing three or more nearest neighbors modes measurements is one of the solutions. We could think as well about relaxing the nearest modes requirement and investigating the situation with non-local coincidence distributions. Then more advanced techniques from the theory of spin glass could be applied, see e.g., \cite{Rams,Jalowiecki}. Finally, our function (\ref{goodness}) to be minimized when restricted to pairs of binary modes is equivalent to the Hamiltonian $H=\sum_{i>j}H_{i,j}$, where
\begin{equation}
H_{i,j}=h_{i,j}{\bf 1}_{i,j}+h'_{i,j}\sigma^z_i\sigma^z_{j}+h''_{i,j}\sigma_i^z\otimes{\bf 1}_{j} +h'''_{i,j}{\bf 1}_{i}\otimes\sigma^z_j.
\end{equation}
Here $\sigma^z_i$ is the Pauli $z$ matrix in mode $i$. The coefficients can be chosen to correspond to $r_{n_i,n_j}$ from (\ref{goodness}). The ground state of Hamiltonian $H$ can be found efficiently by simulated or quantum annealing under some conditions about spectrum of $H$ and the correspondence of $H$ to the connections in the annealer \cite{Kirkpatrick,Kadowaki,Lucas}. In our approach we have some freedom in choosing pairs of modes to guarantee that these conditions are satisfied. So, the simulated or quantum annealing could be used to extend our approach. 

Our technique can be applied to marginal distributions measured in realistic experiments or calculated assuming uniform losses \cite{Garcia,Oszmaniec}. The joint distribution after losses may be interpreted as a lossless distribution multiplied by a stochastic matrix $L$ describing the process of losses, where $L$ is in the form of a tensor product. The measurement matrix from our technique is now the product of $A$ defined as previously and $L$. The new measurement matrix inherits features allowing for keeping the complexity tractable. The impact of losses may be a direction of further studies.

\acknowledgments
We would like to thank Nicol{\'a}s Quesada, Ra{\'u}l Garc{\'i}a-Patr{\'o}n and Jonathan Dowling for constructive comments.
This work was supported by JST CREST Grant No. JPMJCR1772.

\end{document}